# EARLY STAR CHARTS OF THE DUTCH EAST INDIA COMPANY

Richard de Grijs

*School of Mathematical and Physical Sciences, Macquarie University,
Balaclava Road, Sydney, NSW 2109, Australia*
Email: richard.de-grijs@mq.edu.au

**Abstract:** As the European maritime powers expanded their reach beyond north Atlantic coastal waters to distant lands as far away as the East Indies, access to a practical means of maritime navigation in the southern hemisphere became imperative. The first few voyages undertaken by the Dutch East India Company and its predecessor explicitly aimed at compiling star charts and constellations that were only visible south of the Equator, as practical navigation aids. The oldest known star atlas of southern constellations was published in 1603 by Frederick de Houtman. Controversies have plagued de Houtman's astronomical credentials from their inception, however, with contemporaries variously attributing the early southern star charts to Pieter Dirkszoon Keyser, de Houtman, or even to their tutor Petrus Plancius. The balance of available evidence suggests that Keyser initially led the astronomical observing campaign, ably assisted by de Houtman. Upon Keyser's untimely death, de Houtman embraced a leading role in compiling astronomical observations for maritime navigation purposes, whereas Plancius most probably led the delineation of the 12 new southern constellations that soon became part and parcel of the nautical consciousness.

## PREAMBLE

Over the course of the past decade, I have become increasingly interested in exploring the history of maritime navigation, specifically from an astronomical perspective. This has resulted in the publication of a monograph on the determination of longitude at sea in the 17$^{th}$ century (de Grijs, 2017), followed by a few dozen peer-reviewed and popular articles on aspects of the perennial 'longitude problem' that affected oceanic shipping until well into the 19$^{th}$ century.

Upon my relocation to Australia in early 2018, I was keen to expand my history of astronomy-related scholarship to include regional aspects. An initial series of two research articles focusing on science on the 'First Fleet' resulted (de Grijs and Jacob 2021a,b). The First Fleet sailed from England to Australia in 1787–1788, and our research focused, in particular, on the establishment of the first astronomical observatory in New South Wales by second lieutenant William Dawes of the Marines.

While researching the history of longitude determination and the associated longitude awards offered by the authorities in the Dutch Republic in the 16$^{th}$ to 18$^{th}$ centuries (de Grijs, 2021), I came across one of Wayne Orchiston's articles that addressed the history of Indonesian astronomy. During the period of interest, Dutch and Indonesian histories were closely intertwined, although not necessarily happily so at all times. I was particularly intrigued by the following passage from that article:

> In 1917 the distinguished British amateur astronomer, Edward Ball Knobel (1841–1930, […]) published a paper on [Frederick] de Houtman and his star catalog in *MNRAS* (Knobel, 1917[a]), but since then no-one has made an in-depth study of this important astronomer and his pioneering star catalog (though see Dekker, 1987).

> Here is an exciting project for an Indonesian astronomer: the relevant records are in Holland patiently awaiting your attention! (Orchiston, 2017: 147).

Wayne's encouragement inspired me to look into the early star charts of the southern sky produced by astronomers associated with the Dutch East India Company and its precursor. You are currently reading the resulting review article. Despite Wayne's assertion that no-one had made an in-depth study of de Houtman's work, I came across precisely such a study by Verbunt and van Gent (2011).[1] Nevertheless, most studies of this kind tend to focus on relatively narrowly constrained aspects of the history of maritime navigation. In the present article, I have therefore attempted to offer a more comprehensive overview of how de Houtman and his contemporaries went about obtaining their observations, the trials and tribulations they encountered during their voyages, and a characterisation of the resulting star charts and catalogues.

I have expressly refrained from composing an overall biography of de Houtman and of the context in which he lived his life. The interested reader is instead referred to the recent comprehensive account, *Spice at Any Price*, by Howard Gray (2019).

## 1 THE *EERSTE SCHIPVAERT* TO THE EAST INDIES

In the Dutch Republic, the Age of Discovery encompassed the nation's 'Golden Age' (approximately 1588–1672), which led to a significant expansion of trade activity well beyond the relatively safe waters of the North Sea and the Hanseatic ports of northern Europe to destinations as far away as the East Indies, present-day Indonesia. This required advanced navigational aids to secure safe travel across the open oceans (Schilder and van Egmond, 2007; de Grijs, 2021).

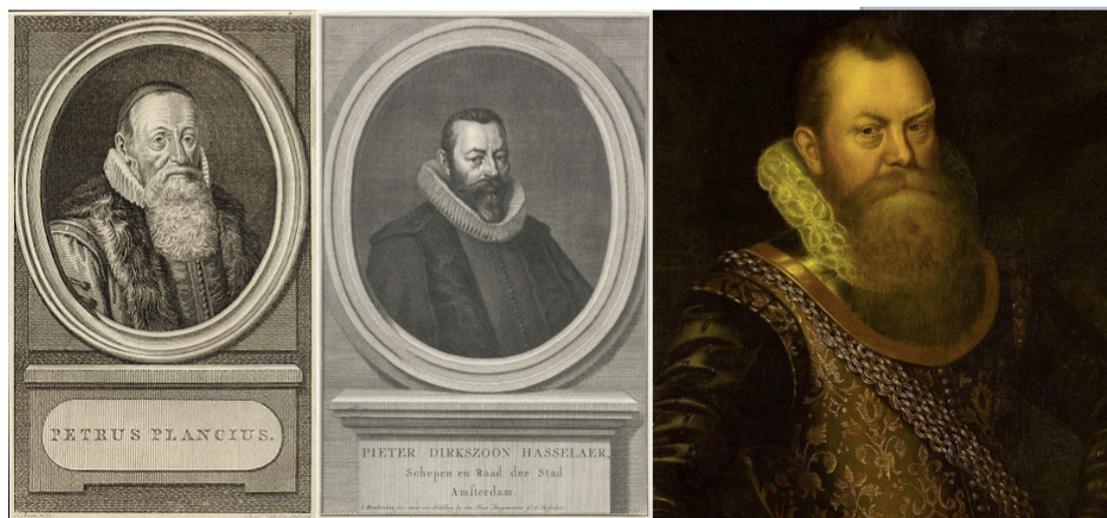

**Figure 1.** Principal characters driving the narrative in the present article. (*left*) Petrus Plancius (1552–1622). (Wikimedia Commons; public domain). (*middle*) Pieter Dirkszoon Keyser (ca. 1540–1596). (Google Arts and Culture; public domain). (*right*) Frederick de Houtman (1570/1–1627). (Rijksmuseum, SK-A-2727, via Wikimedia Commons; Creative Commons CC0 1.0 Universal Public Domain Dedication).

The joint pursuit of trade and practical science came naturally to the sailors engaged in the Dutch East India voyages. Scientific endeavours were pursued systematically ever since the first Dutch voyage to Asia of 1595–1597, commonly known as the *Eerste Schipvaert* (van Berkel, 1998). On that voyage, Petrus Plancius (1552–1622; Figure 1, left)—the Dutch–Flemish astronomer, cartographer and theologian—ordered that sufficient numbers of observations of variations of the 'magnetic declination' (deviations of the compass needle from true North) be obtained as a potential but ultimately unsuccessful means to determine one's longitude at sea (de Grijs, 2021).

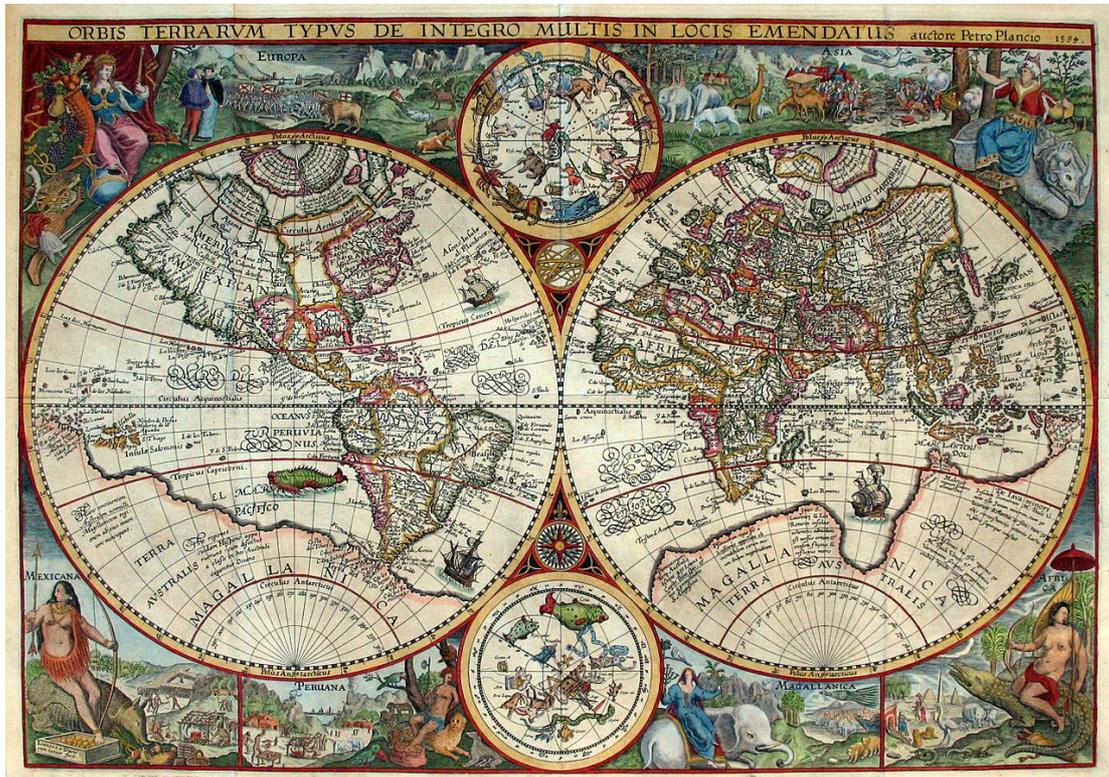

**Figure 2.** *Orbis terrarum typus de integro multis in locis emendatus Petro Plancio* (Whole world map with many additions by Petrus Plancius, 1594). (Wikimedia Commons; public domain).

Plancius produced the earliest-known maps of the southern sky. An extremely rare Plancius map from 1592 (Blundeville, 1636; Knobel, 1917b) preceded a more commonly available version. The latter, published in 1594, is included in the *Orbis terrarum typus de integro multis in locis emendatus Petro Plancio* (Whole world map with many additions by Petrus Plancius, 1594; see van Linschoten, 1599). Plancius' 1594 map (Figure 2) includes the 48 Ptolemaic constellations published in Ptolemy's masterwork *Mathēmatikē Syntaxi*, better known as the *Almagest* ($2^{nd}$ century CE), as well as Columba (whose stars were known by Ptolemaic scholars), Crux (also known by Ptolemy, but here listed as a separate constellation for the first time), Eridanus (expanded from Ptolemy's $34^{th}$ star to α Eridani), Triangulum Australe and a large constellation in the shape of a man known as 'Polophilax' (Knobel, 1917a). Scholars agree that at this time, no other southern hemisphere constellations were known, a notion supported by Thomas Hood's (1556–1620) statement associated with his *Celestial Map* of 1590 (Figure 3) that no stars other than those listed by Ptolemy had been observed (Knobel, 1917a).

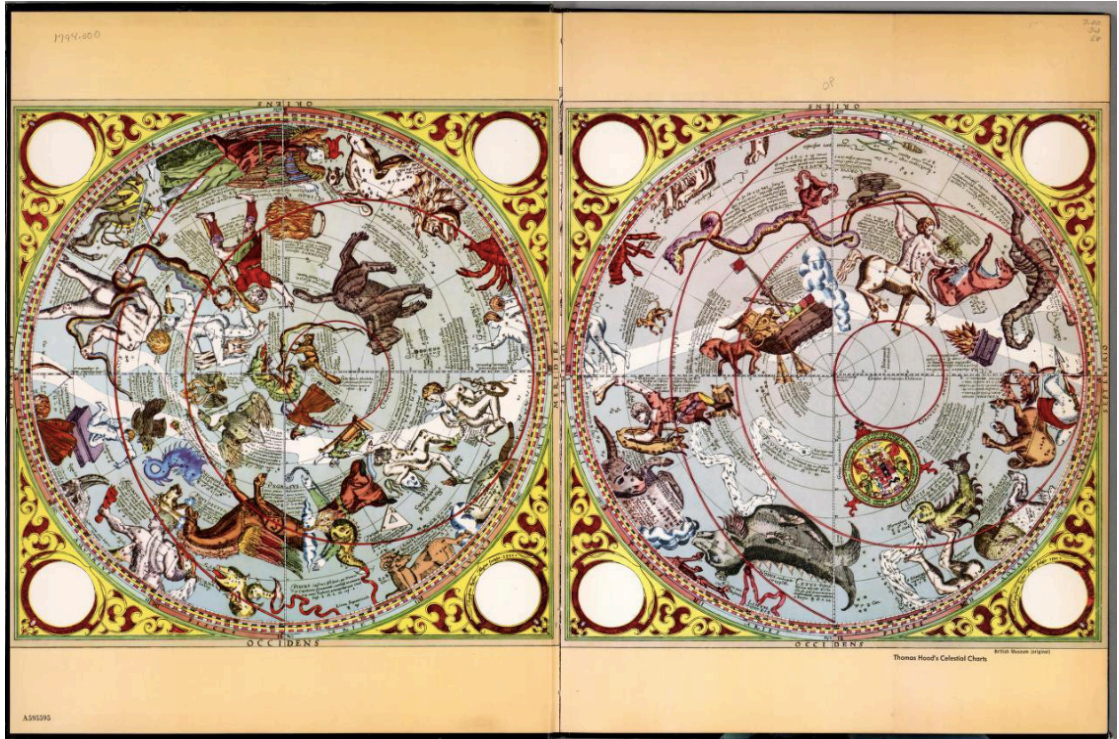

**Figure 3.** Thomas Hood's *Celestial Map*, 1590. (© David Rumsey Map Collection; reproduced with permission).

Therefore, when the *Eerste Schipvaert* left the Dutch East India Company's Texel roads on 2 April 1595, Plancius instructed his disciple Pieter Dirkszoon Keyser van Em(b)den (ca. 1540–1596; Figure 1, middle)—also known as Petrus Theodorus F(ilius) Embdanus or Peter Theodors Sohn (e.g., Moll, 1825; Knobel, 1917a)—to obtain observations of the stars around the South Celestial Pole. The *Eerste Schipvaert* expedition consisted of four ships, including the *Hollandia* (or *Hollandsche Leeuw*, 'Dutch Lion'), *Mauritius*, *Amsterdam* and *Het Duyfken* ('the Little Dove'). Keyser was principal navigator on the *Hollandia*, and subsequently on the *Mauritius*, starting as head of the steersmen on the former, under captain Jan Dignum(s)z (d. 1595).

The *Hollandia* arrived at the Baie de Saint-Augustin in Madagascar on 2 or 3 September 1595, desperate to obtain fresh supplies and recover from the hardships endured *en route*, including numerous instances of scurvy. The convoy remained anchored and wind-bound at Madagascar for several months, until April or May 1596. During this time, Keyser (who was well versed in mathematics and astronomy)

> … sought comfort in science, and enriched his knowledge of astronomy by improving the position of old and the observation of new constellations. (Knobel, 1917a: 417).

Knobel (1917a) further suggests that at the latitude of Madagascar, approximately 47° South, he would have been able to observe stars as faint as $5^{th}$ or $6^{th}$ magnitude in the South Celestial Pole region. Knobel then adds that Keyser most likely used Tycho Brahe's (1546–1601) observations, which were supplied by Plancius, to determine the approximate right ascensions of the newly observed stars.

Paulus Merula (1558–1607), the historian and geographer, explained that Keyser made his observations from the ship's crow's nest using an unspecified instrument he had received from Plancius (Merula, 1605). That instrument was most likely a cross-staff, a quadrant or an *astrolabium catholicum*, that is, a universal astrolabe, which represented an innovation with respect to the planispheric astrolabe in common use in that it would work at any latitude.

The fleet comprising the *Eerste Schipvaert* eventually arrived in the Sunda Strait and at Bantam (now Banten, in West Java, present-day Indonesia) in September 1596. Keyser died soon after their arrival, some time between 11 and 13 September 1596 (ab Utrecht Dresselhuis, 1841; de Waard, 1912: 674). Keyser was clearly held in high regard by the Company of Distant Lands, the predecessor of the Dutch East India Company, which referred to him as "… a highly experienced sailor, by whose death the Company of Distant Lands—currently at Bantam—has lost a lot." (ab Utrecht Dresselhuis, 1841: 530; own translation).

Keyser's observations reached Plancius following the fleet's return to the Texel anchorage on 10 August 1597.

## 2 CONTROVERSIAL CREDIT

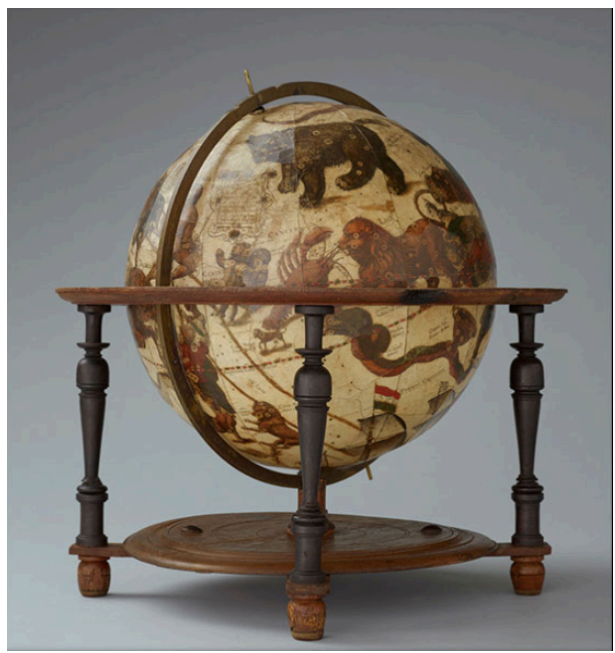

**Figure 4.** Celestial globe made by Jodocus Hondius, from 1600. (© Linda Hall Library; reproduced with permission).

The astronomical observations and the resulting constellations derived from the voyage of the *Eerste Schipvaert* were first published in 1597 or 1598 on a celestial globe produced by Plancius (Dekker, 1987), and again a year or two later on a globe made by the Dutch cartographer Jodocus Hondius (1563–1612; Figure 4; van der Krogt, 1993). Although some of the stars included among Keyser's observations had been known previously (e.g., 107 of Keyser's stars were part of the Ptolemaic constellations; Knobel, 1917a; Tichelaar, n.d.), delineation of 12 new southern constellations is usually credited to Plancius, Keyser and Frederick de Houtman (1570/1–1627; Figure 1, right).

Unfortunately, Keyser's original notes have been lost, and so we do not know much else about the man's life or accomplishments. It is, therefore, not clear whether the definition of those 12 new southern constellations is entirely attributable to Keyser. Dekker (1987) has suggested that Plancius himself stood at the basis of the new constellations (see also Verbunt and van Gent, 2011; Tichelaar, n.d.), whereas the scholarly and popular literature is rife with suggestions that it was, instead, de Houtman who should receive full credit.

Let us consider this latter claim in more detail. De Houtman was the younger brother of Cornelis de Houtman (1565–1599), the merchant seaman in overall command of the *Eerste Schipvaert*. Frederick de Houtman sailed as volunteer sub-commissioner on the *Hollandia*, supporting the expedition's mercantile aspects. He is thought to have assisted Keyser with his astronomical observations and likely made independent astronomical observations himself, including substantive contributions to the 12 newly delineated constellations—although his astronomical credentials remain fiercely debated (e.g., ab Utrecht Dresselhuis, 1841; Knobel, 1917a; Van Lohuizen, 1966; van der Sijs, 2000; Verbunt and van Gent, 2011; Ridpath, 2018: Ch. 1, p. 3; Tichelaar, n.d.; and references therein).

Frederick himself states in the *Introduction* to the catalogue of southern stars he eventually published (see Section 4) that he made some observations himself during the voyage of the *Eerste Schipvaert*, *viz*. "Also added [are] the declination of several fixed stars which during the [*Eerste Schipvaert*] I have observed around the South Pole." (de Houtman, 1603: *Introduction*; own translation). De Houtman's assertion is supported by statements from both the Dutch geometer and astronomer Adriaan Adriaanszoon (1571–1635)—better known as Metius ('measurer')—and the well-known cartographer Willem Jansz. Blaeu (1571–1638), against the objections of Merula (ab Utrecht Dresselhuis, 1841). Metius (1621: 4–5) explicitly endorsed de Houtman's credentials:

> Near the South Pole there are numerous stars that do not appear above our horizon: of which the most important ones have been carefully observed by the audacious Governor Frederick [de] Houtman (**at one time my disciple in astronomy**) in the East Indies on the island of Sumatra, … (own translation; my emphasis).

On the other hand, it is plausible that de Houtman had unfettered access to Keyser's observations on the long voyage home, following the latter's demise in Sumatra, and so it is not inconceivable that de Houtman may have tried his hand at grouping the newly observed stars into easily recognisable constellations.

Verbunt and van Gent (2011) undertook a careful, quantitative comparison of the early modern star catalogues of the southern sky published by de Houtman in 1603 (see Section 4), Kepler in 1627 (specifically the *Rudolphine Tables*, with the addition of stars he referred to as second and third class, *Secunda classis* and *Tertia classis*) and Halley in 1679. They concluded that the observations used to delineate the new constellations were mostly obtained during the *Eerste Schipvaert* (see also Stein, 1917). This agrees with de Houtman's statement in his *Introduction*, although he does not acknowledge any contribution by Keyser. Verbunt and van Gent (2011) further dismiss the suggestion by Knobel (1917a) that de Houtman may simply have plagiarised Keyser's observations, given the impossibility to observe many of the catalogued stars from Sumatra (as stated in de Houtman's *Introduction*).

Finally, Verbunt and van Gent (2011) emphasise that Dekker (1987) found significant differences between the new constellations as represented on the Hondius globes of 1598 and 1601 (which had to be based on observations obtained during the *Eerste Schipvaert*) and those on Blaeu's globe of 1603 (Figure 5). The latter, which included stars that expanded the Ptolemaic constellations, relied on additional observations obtained by de Houtman during his second voyage to the East Indies of 1598–1602 (see Section 3). Contrary to Dekker's (1987) assertion that Keyser and de Houtman may have obtained independent observations, Verbunt and van Gent (2011) suggest that Plancius and Blaeu may instead have independently *analysed* a common data set.

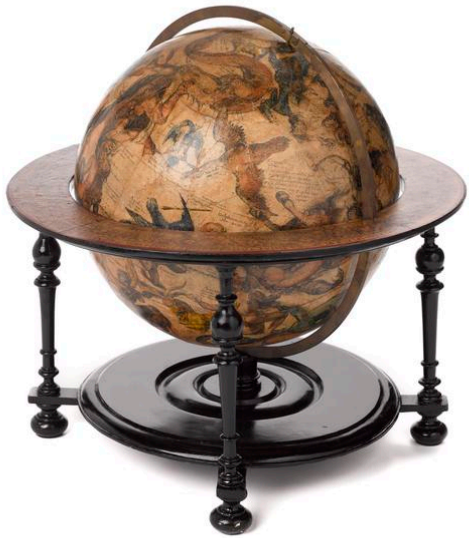

**Figure 5.** Celestial globe made by Willem Janszoon Blaeu, from 1603. (© The Board of Trustees of the Science Museum, London; reproduced with permission).

As such, and given the available documentation, it seems most likely that Keyser was the primary observer, whereas de Houtman contributed his share to the final data set. This conclusion is also reflected by Bodel Nijenhuis (1831: 321):

Our [de] Houtman can just as well as P[ieter] Dirkz[oon Keyser] have had the Amsterdammer P[etrus] Plancius as tutor in mathematics. However, what would have been the correct relationship between both astronomers, the Ostfrisian [Keyser] and the Hollander [de Houtman], is as yet not fully clear to me. (own translation).

## 3 FREDERICK DE HOUTMAN'S SECOND VOYAGE

In 1597, on his way home following the *Eerste Schipvaert*, de Houtman may have started planning a second observing campaign of the southern sky already. His next voyage to the East Indies would commence shortly. Cornelis and Frederick de Houtman departed on 25 March 1598 once again to the East.

This time, Frederick was captain of the *Leeuwinne* ('Lioness'), whereas the expedition's overall command resided with his brother Cornelis once again. The voyage was ill-fated, however. On 11 September 1599, a day prior to their departure from Sumatra to Johor (Malaysia), 29 of the ships' crew were murdered, including Cornelis de Houtman, whereas Frederick and several tens of their compatriots were taken prisoner by sultan Alauddin Ri'ayat Syah Sayyid al-Mukammal (d. 1605) of Aceh (northern Sumatra).

Frederick remained the sultan's unwilling guest for 26 months. He recorded his experiences, the pressure he was put under to convert and the ordeals he was forced to endure in his prison cell in his *Cort Verhael* ('Brief account'; 1601). Perhaps surprisingly, his tone is more curious and understanding than bitter. It took an intervention by Prince Maurits (Maurice) of Nassau, the Dutch regent, including a shipment of arms, mirrors and money, to secure de Houtman's release in 1601. Meanwhile, Frederick made good use of his time in captivity by studying Malay (the *lingua franca* in Southeast Asia at the time) and by making astronomical observations.

The astronomical observations obtained during his second voyage to the East Indies and his time in captivity supplemented and complemented those made on the first expedition, that is, the eventual catalogue contained a significantly increased number of observations of improved quality, many based on better calculations (e.g., Tichelaar, n.d.). In particular, he improved and expanded the body of observations of the Ptolemaean constellations during his stay in Aceh (Verbunt and van Gent, 2011).

Upon his release, de Houtman returned to Alkmaar in the Dutch Republic in 1602. Settled once again in his hometown, he completed a Dutch–Malay–Malagasy dictionary and grammar guide, published in 1603.[2] Curiously, he published the delineations of a set of southern constellations as well as his astronomical observations, both those resulting from the *Eerste Schipvaert* and those from his second expedition, as an appendix to his language compendium. This rather unusual choice resulted in the catalogue's initial decline into obscurity. De Houtman's catalogue, the oldest surviving catalogue of southern stars published in book form, contained "… many fixed stars, located around the South Pole, never seen before this time." (de Houtman, 1603: *Introduction*). He explicitly declared that it was based on his own observations,

> There will be found at the end the declinations of several fixed stars in the region of the South Pole which I had observed on my first voyage, and which on my second voyage I revised and corrected with more care and brought up to the number of 300, as may be seen on the Celestial Globe published by William Jansen [Blaeu]. (de Houtman, 1603: Dedicatory letter to the vocabulary; transl.: Knobel, 1917a).

… and that it aimed to "serve all sailors, who navigate south of the equinoctial line and are of interest to all lovers of astronomy or the mathematical arts." (de Houtman, 1603: *Introduction*; transl.: Knobel, 1917a).

De Houtman obtained a formal privilege for his publication from the States General of the United Provinces of the Dutch Republic (the national government) for eight years; an extract from the official award follows (own translation):

> On the basis of the patent letters, the States General of the United Dutch Provinces have awarded their honorable servant Frederik Pietersz. de Houtman the sole privilege for the next eight years to have printed, published and sold a certain Language Book or Dictionarium of Dutch, Malay and Malagasy, together with many Turkish and Arabic words, and moreover a few fixed stars that are close to the South Pole, to about thirty-five degrees South of the Equator, about three hundred in number. They have prohibited that this Language Book bearing the aforesaid stars be reprinted, in whole or in part, within these United Countries during the aforementioned eight-year period, or that it, after reprinting elsewhere, be imported into such Countries under any pretext, under penalty of forfeiture of the reprinted copies, and in addition of the sum of one hundred Carolus guilders, to be spent as follows: one third for the Officer, one third for the poor and the remaining one-third for the aforementioned Frederik de Houtman, as is more fully apparent from the said patent letters.
>
> Dated 4 February 1603. Sealed and signed by settlement of the States General.
> C. Aerssens[3]

# 4 DE HOUTMAN'S SOUTHERN STAR CATALOGUE

As a practical guide, the book was used extensively during subsequent expeditions. It was popular among merchants and interested laypeople alike, and so it was republished and translated a number of times. In both 1673 and 1680, the language compendium was reissued by order of the Governors of the Dutch East India Company (de Houtman, 1673–1680). The star catalogue and the Dutch–Malagasy dictionary had been omitted, the latter presumably because South Africa rather than Madagascar had become the usual way station. The dictionary component was followed by twelve practical dialogues in Malay, which took centre stage over the three Malagasy dialogues from the first edition (van der Sijs, 2000; Tichelaar, n.d.) and for which de Houtman had presumably collected his information during their extended sojourn in Madagascar in 1595–1596 (Knobel, 1917a).

The star catalogue's title page in the appendix of the 1603 edition reads thus:

Here follow several fixed stars [observed], with efficient instruments, by Frederick de Houtman, in the island of Sumatra, [their positions] corrected and their numbers increased. For the use and service of those who navigate South of the equinoctial line, also for all amateurs and those who have occasion for the best. These stars are arranged according to their Right Ascension; that is, the degree and minute which a star in the South or North has from where the equinoctial line cuts through [sic]. Declination is the number of degrees and minutes a star is distant from the equinoctial line towards the South or North Pole. Magnitude is the size of the stars : often a star is of the first size or greatest light : thus there are seven [sic] degrees of size and light. (translation: Knobel, 1917a).

It contained observations of 304 stars overall, although for one star (located in the tail of the constellation Scorpius) celestial coordinates are lacking. Of the remaining 303 stars, which were only visible from the southern hemisphere, 107 were already known by Ptolemy (Knobel, 1917a), whereas for 135 observations had been obtained by Keyser. Since Keyser's original notes have been lost, it is unclear whether the remaining 61 stars were, in fact, observed by him or if they represent de Houtman's unique contributions. Careful comparison of Keyser's and de Houtman's positions reveals numerous discrepancies, and so it would not be out of character to consider de Houtman as the principal observer of the latter stars (e.g., Verbunt and van Gent, 2011; Tichelaar, n.d.). In any case, de Houtman's (1603) southern star catalogue provided the basis for the renaming of many of the southern constellations (Hidayat, 2000). It was translated into French in 1881.

The 303 stars observed by Keyser and de Houtman found their way into 21 constellations, among which 12 were new: see Table 1. Of the remaining nine, eight constellations were already known by Ptolemy (see Table 2), whereas the ninth is Cruzeiro (De Cruzero), the Spanish or Southern Cross, for the first time separated from Centaurus and accurately depicted. Of the 303 stars in de Houtman's catalogue, 111 were members of the 12 newly delineated constellations; the majority were, however, previously known or fainter members of the Ptolemaic constellations, including 56 stars in Argo Navis (comprising the modern constellations Carina, Puppis and Vela), 48 in Centaurus, as well as stars in Ara, Corona Australis, Crux, Lupus, Columba,[4] Scorpius and southern Eridanus, which de Houtman called 'den Nyli', that is, the Nile (Ridpath, 2018).

**Table 1: New southern constellations following Plancius, Keyser and de Houtman.**

| De Houtman's designation | English translation | Modern designation | # stars (de Houtman) |
|---|---|---|---|
| Den voghel Fenicx | The Phoenix bird | Phoenix | 13 |
| De Waterslang | The water snake | Hydrus | 15 |
| Den Dorado | | Dorado | 4 |
| De Vlieghe | The fly | Musca | 4 |
| De vlieghende Visch | The flying fish | Volans | 5 |
| Den Camelion | The chamaeleon | Chamaeleon | 9 |
| Den Zuyder Trianghel | The southern triangle | Triangulum Australe | 4 |
| De Paradijs Voghel | The bird of paradise | Apus | 9 |
| Den Pauw | The peacock | Pavo | 19 |
| Den Indiaen | The Indian | Indus | 11 |
| Den Reygher | The heron | Grus | 12 |
| Den Indiaenschen Exster, op Indies Lang ghenaemt | The Indian magpie, named Lang in the Indies | Tucana | 6 |

**Table 2: Known southern constellations observed by Keyser and de Houtman.**

| De Houtman's designation | English translation | Modern designation | # stars (de Houtman) |
|---|---|---|---|
| De Duyve met den Olijftack | The dove with the olive branch | Columba | 11 |
| De Zuyder Kroon | The southern crown | Corona Australis | 16 |
| Het Zuyder Eynde van de Nyli | The southern end of the Nile | Eridanus (southern section) | 7 |
| Argo Navis | The ship (Argo) | Carina, Puppis and Vela | 56 |
| Centaurus | Centaur | Centaur | 48 |
| De Cruze(i)ro | The southern cross | Crux | 5 |
| Lupus, den Wolf | The wolf | Lupus | 29 |
| Het Outaer (Altaar) | The altar | Ara | 12 |
| Den steert van Scorpio | The tail of the scorpion | Scorpio (tail) | 9 |

## 5 KEYSER OR DE HOUTMAN?

Following the earliest depictions of the new constellations on the globes of Plancius and Hondius of 1597/8–1601, adjustments based on de Houtman's updated observations and calculations soon found their way onto next-generation globes, such as Blaeu's Celestial Globe of 1603 (Dekker, 1987). The latter saw the light of day prior to the publication of de Houtman's catalogue and approximately at the same time as the German astronomer Johann(es) Bayer's (1572–1625) *Uranometria Omnium Asterismorum* (Uranometry—celestial cartography—of all the asterisms; Augsburg, September 1603). Bayer's 49th map in his *Uranometria* star atlas (Figure 6) shows all 12 new constellations. In addition,

> His map of Eridanus shows the continuation beyond the last of Ptolemy's stars to α Eridani; it also gives some stars in Phoenix and Doradus; that of Canis Major gives Columba, "recentioribus Columba"; Argo shows some stars in Volans; Centaurus includes Crux as a figure, and Triangulum Australe; Ara shows some stars in Triangulum, Grus, and Pavo; Piscis Austrinus includes stars in Grus. (Knobel, 1917a: 414).

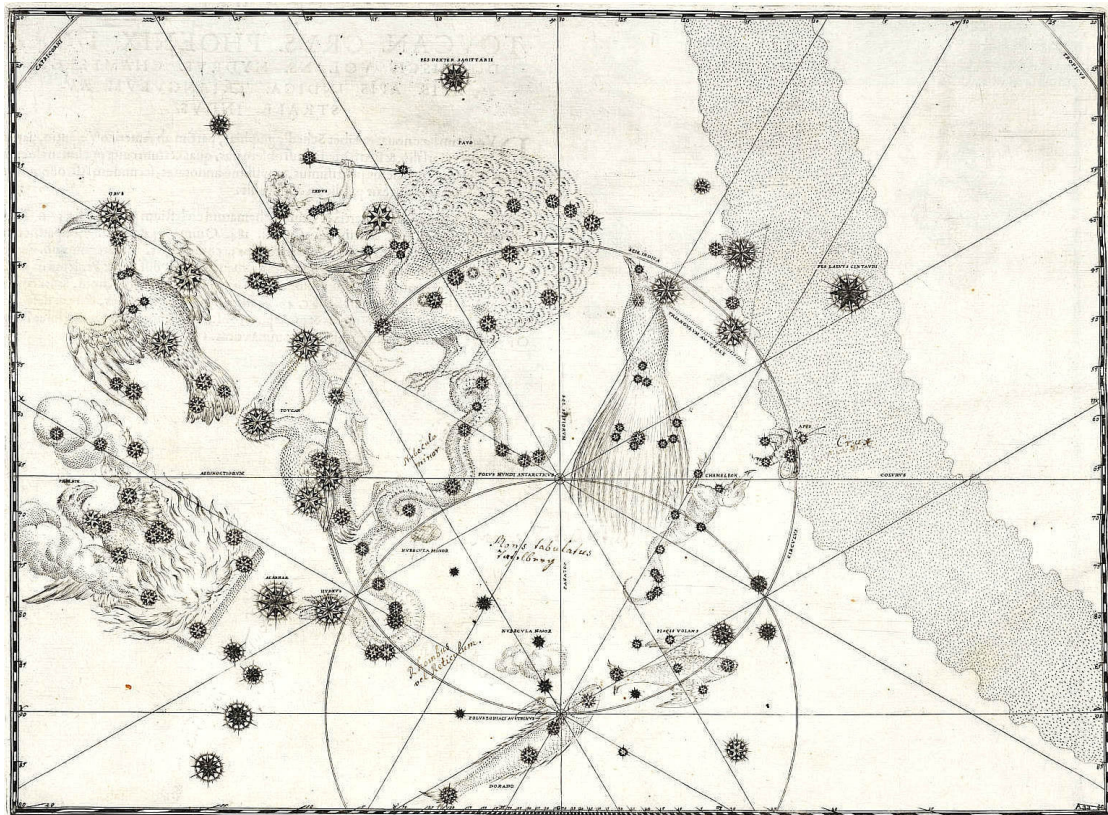

**Figure 6.** 49th map from Johann Bayer's *Uranometria*, showing the new southern constellations (© Linda Hall Library; reproduced with permission).

Bayer's accompanying text clarifies that the stars included in this 49th map were observed, in part, by Amerigo Vespucci (1451–1512), Andrea Corsali (b. 1487) and Pedro de Medina (1493–1567; de Grijs, 2020), but that their positions were derived by the "most learned" seaman Petrus Theodorus, that is, by Keyser. Corsali, in particular, deserves more than a cursory mention here, given that the Italian explorer was the first European to describe, identify and record the five stars of Crux, the "marveylous crosse" in 1516: "The crosse is so fayre and bewtiful that none other heavenly signe may be compared to it." (Power, 2018; Figure 7).

Blaeu was, however, clearly enamoured by de Houtman's work. In the text accompanying his Celestial Globe he wrote,

In the section of the heavens which borders the South Pole, F. H. [Frederick de Houtman], then on the island of Sumatra, measured many stars, and formed thirteen [sic] constellations of them. (van der Aa, 1867: 1331; own translation).

… while on his globe we read, in Latin,

We have added more than 300 stars to the South – and the for us always hidden – Pole. Their distances from the stars determined and known by Tycho [Brahe], were measured by F.H. [Frederick de Houtman] who separated them into constellations, of which we have traced all positions on this globe back to the year 1640. (van der Aa, 1867: 1331; own translation from Dutch).

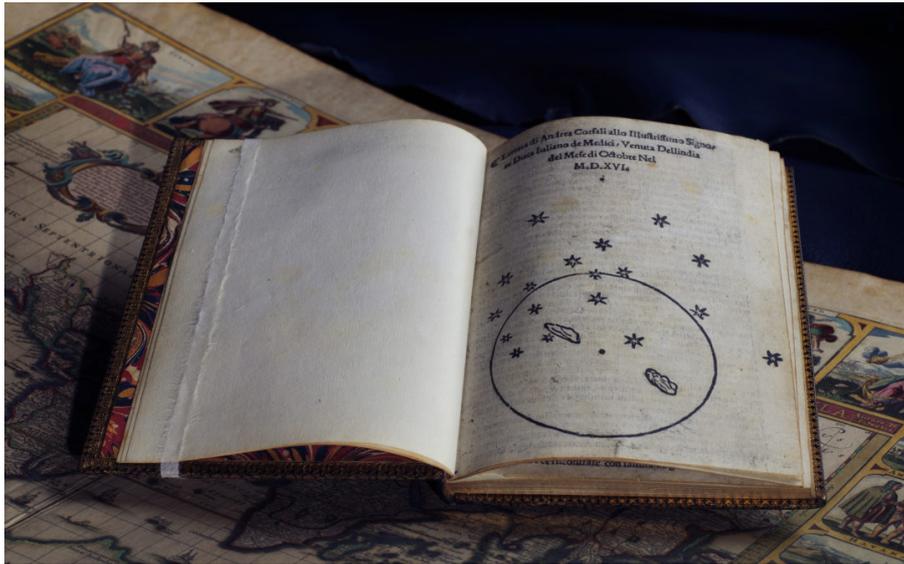

**Figure 7.** Andrea Corsali's original depiction of the Southern Cross. (© State Library of New South Wales; reproduced with permission).

The provenance of the constellations on Blaeu's Celestial Globe of 1603 has been the subject of debate since its inception. Whereas Blaeu credited de Houtman as his source, Merula stated in his *Cosmographia Generalis* (1605) that all observations of the "longitudes, latitudes, declinations, etc." were obtained by Petrus Theodorus (Keyser). On the other hand, the German lawyer Julius Schiller (ca. 1580–1627) published a star atlas, *Coelum Stellatum Christianum* (Augsburg, 1627), in which he followed Blaeu and noted that his constellations are found on globes by "Petrus Plancius, or Petrus Karif, Jansonius (Blaeu) and [de] Houtman, etc." (Knobel, 1917a: 415). Also in 1627, Johannes Kepler (1571–1630) released a newly edited version of Brahe's tables as part of his *Rudolphine Tables*, to which he appended two additional catalogues with stars belonging to the *Secunda classis* and *Tertia classis* (second and third class, respectively):

> *The third class* of fixed stars[:] comprising twelve celestial images, which can not be seen at all in our moderate northern zone. In his *Uranometria* Joh. Bayer reports that these have been observed by Amerigo Vespucci, Andreas Corsali and Pedro de Medina, the first among Europeans, and declares that they were for the first time corrected to astronomical standard by **Pieter Dicksz. [Keyser]**. Jacobus Bartsch from Lausitz, a diligent young man, famous for some time now for his great merits concerning the celestial globe, assembled these same [constellations] into numbers and a map from the last tables and manuscripts of Johann Bayer himself (a splendid little collection of Christian constellations extracted from the *Uranographia* of Schiller, the publication of which is forthcoming in accordance with the last will of the author); and he has promised that he will subsequently publish the most perfect maps, by producing a one-and-a-half foot globe with the ancient images, as more conform with the version of Tycho. (Verbunt and van Gent, 2011: 2; my emphasis).

Finally, Edward Sherburne (1618–1702), the English poet, published an astronomical appendix associated with his translation of the poem *The Sphere of Marcus Manilius* (1675), stating that the constellations discussed in that poem were

> … first found out and denominated by some eminent navigators sayling beyond the line, as particularly by Americus Vespuccius, Andreas Corsalius, Petrus [de] Medina, but principally by Fredericus [de] Houthman, who, during his abode in the island of Sumatra, made exact observations of them, being by **Petrus Theodorus [Keyser]** and Jacobus Bartschius reduced into order. (Knobel, 1917a: 415; my emphasis).

Nevertheless, de Houtman's purported accomplishments were soon recognised more widely. As a case in point, Benjamin Apthorp Gould (1824–1896) included de Houtman's stars in his *Uranometria Argentina* of 1879 (Knobel, 1917a), the leading star atlas of the day. François Valentijn (1666–1727), best known for his seminal work *Oud en Nieuw Oost-Indiën* ('Old and New East India'; 1724), highlighted the astronomer's achievements:

> He has made several fine astronomical observations [while imprisoned] and discovered a number of hitherto unknown stars, which he subsequently also distributed in print. (Valentijn, 1724: 172–174; own translation).

Verbunt and van Gent (2011) similarly concluded that de Houtman had made highly accurate measurements. However, among his approximate contemporaries, one high-profile natural philosopher was less enamoured by de Houtman's accomplishments. Edmond Halley (1656–1742) compiled his own southern star catalogue from St Helena in 1676–1677, and commented scathingly on de Houtman's earlier work:

> Then there is a rumour that a certain Dutchman Frederick [de] Houtman has made an effort on these stars on the island of Sumatra, and that Willem Blaeu [used] his observations to correct the celestial globe which he [Blaeu] published. Which instruments he used is not known to me, but from a comparison made of his globe with our catalogue, it is sufficiently and abundantly clear that this observer was little practised in this arena. (Halley, 1679; transl.: Verbunt and van Gent, 2011: 11).

Of course, Halley's instrumentation was far superior to that used by de Houtman. The former had access to a state-of-the-art astronomical sextant with telescopic sights, which had been made specifically for his observations in St Helena, probably by the Ordnance Office (Cook, 1998: 38).

On the balance of all available evidence, and following more recent quantitative analyses, it appears that de Houtman's alleged contributions to the delineation of 12 new southern constellations—and by extension to improved navigation practices at sea—cannot be simply dismissed. He was clearly an educated man, well versed in mathematics and astronomy, and he had numerous opportunities to make careful astronomical observations. The star catalogue resulting from his first two voyages to the East was most probably a result of the combined efforts of Keyser and de Houtman, whereas Plancius likely also played a significant role in the definition of the new constellations. Much of the controversy surrounding de Houtman's astronomical credentials is likely driven by his failure to credit Keyser's contributions following the latter's untimely death.

# 6 NOTES

[1] While perusing Verbunt and van Gent (2011), I noticed an issue that requires correction. For the record, these authors refer to de Houtman's entry F46 in their Table 5, Section 5.4 and Appendix B.1 as a star in the Small Magellanic Cloud (which is correct), whereas in their Section 5.1 they incorrectly classify the same entry as belonging to the Large Magellanic Cloud.

[2] The book contained four lists of words (Tichelaar, n.d.), including Dutch–Malay (2638 words), Dutch–Malagasy (2505 words), Dutch–Turkish (1098 words) and Dutch–Arabic (1096 words).

[3] Cornelis Aerssens was clerk of the States General from 1584 to 1623.

[4] De Houtman referred to this constellation as 'De Duyve met den Olijftak' (the dove with the olive branch).

# 7 ACKNOWLEDGEMENT

In recognition that this contribution will form part of a volume published in celebration of Wayne Orchiston's 80$^{th}$ birthday, this appears an opportune place to acknowledge Wayne's encouragement to explore a range of history of astronomy aspects in more detail than I had initially planned to do. This led directly to my deep engagement with research into the history of science as an additional professional focus area.

**ABOUT THE AUTHOR**

Professor Richard de Grijs obtained his PhD in Astrophysics from the University of Groningen (Netherlands) in 1997. He subsequently held postdoctoral research positions at the University of Virginia (USA) and the University of Cambridge (UK), before being appointed to a permanent post at the University of Sheffield (UK) in 2003. He joined the Kavli Institute for Astronomy and Astrophysics at Peking University (China) in September 2009 as a full Professor. In March 2018, Richard moved to Macquarie University in Sydney (Australia) as Associate Dean (Global Engagement). Richard was a Scientific Editor of *The Astrophysical Journal* since 2006 and took on the role of Deputy Editor of *The Astrophysical Journal Letters* in September 2012. He held this latter role until mid-2018. He has joined the Editorial Team of the *Journal of Astronomical History and Heritage* as an Associate Editor in early 2021. Richard received the 2012 Selby Award for excellence in science from the Australian Academy of Science, a 2013 Visiting Academy Professorship at Leiden University from the Royal Netherlands Academy of Arts and Sciences, a 2017 Erskine Award from the University of Canterbury (New Zealand), as well as a Jan Michalski Award from the Michalski Foundation (Switzerland), also in 2017.  His research focuses on the astronomical distance scale, as well as on many aspects of star cluster physics, from their stellar populations to their dynamics and their use as star-formation tracers in distant galaxies. He is also engaged in a number of research projects related to the history of astronomy, with particular emphasis on maritime navigation in the seventeenth century. In 2017, he published *Time and Time Again: Determination of Longitude at Sea in the 17th Century* (IOP Publishing). On Sundays, Richard can be found at the Australian National Maritime Museum as volunteer guide on any of the Museum's historical vessels, which include replicas of Captain Cook's H.M.B. *Endeavour* and of the Dutch East India Company's yacht *Duyfken*. The latter made the first recorded European landing on Australian soil (Cape York), in 1606.